\documentclass
[preprint,prc,showpacs,showkeys,a4paper,amsfonts,amssymb]{revtex4}%
\usepackage{amsfonts}
\usepackage{amsmath}
\usepackage{amssymb}
\usepackage{graphicx}%
\begin{document}
\preprint{ }
\title{Correlations in Self-Bound \\Systems of Identical Fermions}
\author{G. Kamuntavi\v{c}ius, A. Ma\v{s}alait\.{e}, S. Mickevi\v{c}ius}
\affiliation{Vytautas Magnus University, Donelai\v{c}io 58, LT-44248 Kaunas, Lithuania}
\author{D. Germanas, R. Kalinauskas}
\affiliation{Institute of Physics, Savanori\c{u} 231, LT-02300 Vilnius, Lithuania}
\author{R. \v{Z}emai\v{c}i\={u}nien\.{e}}
\affiliation{\v{S}iauliai University, Vilniaus 88, LT-76285 \v{S}iauliai, Lithuania}

\begin{abstract}
A method for the calculation of translationally invariant wave functions for
systems of identical fermions with arbitrary potential of pair interaction is
developed. It is based on the well-known result that the essential dynamic
part of the Hamiltonian for the system of identical particles is the Reduced
Hamiltonian operator describing relative movement of two particles inside the
system. The eigenfunctions of this operator take into account all correlations
caused by interaction. These eigenfunctions are basic for the construction of
the components (i.e. the functions with a lower degree of antisymmetry) of the
system wave functions. The main problem of this approach appears to be
antisymmetrization of the components. The developed universal algorithm for
antisymmetrization gives a possibility to perform this operation in a simple
way and keep numerical approximations under control.

\end{abstract}
\maketitle

\section{Introduction}

The wave function of the self-bound system in the absence of external fields
must be invariant in respect of translations in space. This invariance of wave
function means that it is dependent only on intrinsic degrees of freedom of
the system. However, the traditional methods for quantum systems description,
such as Shell Model or Hartree-Fock Self-Consistent Field (SCF) method,
produce wave functions dependent on a set of one-particle variables, hence
also on the system center of mass radius-vector. This\ shortage of mentioned
methods is well-known, however, the wave functions dependent on one-particle
variables are very attractive because they allow simple procedure of
antisymmetrization. In some cases this approximation with the Hamiltonian and
the wave functions, both dependent on redundant variables, cannot produce any
serious problems. The moving and exciting in uncontrollable way center of mass
of the system, such as an atom, molecule, or electron gas in solid state, is
not the problem for the system under consideration (i.e., the system of
electrons) but the problem of external fields, keeping these electrons
together to form such a system. However, for nuclei and hadrons the
translational invariance appears to be a real problem. These systems are
essentially self-bound. The external fields, able to dictate behavior of such
a system in normal conditions, are very weak in comparison with interactions
of nucleons or quarks, i.e., the structural elements of nuclei or hadrons.
This invariance of corresponding wave functions creates less investigated
additional correlations between particles.

A quantum system without correlations is a system, whose wave function can be
present as a product of wave functions of subsystems, composing the entire
system. Such factorization of wave function is possible when interactions
among these subsystems are absent or can be described by some central field
forcing these subsystems to compose the system. Such systems do not exist,
this is only a more or less acceptable model simplifying the description. If
particles compose a system, they always are correlated, and first of all due
to interactions among them. However, besides dynamic correlations, there exist
additional correlations, characteristic namely of quantum systems. These are
correlations created by different symmetries of the system. Any symmetry has a
corresponding operator commuting with Hamiltonian. Wave function of the
quantum system must be an eigenfunction of these operators. This introduces
the new quantum numbers characterizing the system, but at the same time
requires linear combinations of mentioned products of subsystem functions,
hence additional correlations.

The first among them is invariance of the system in respect of rotations and
inversion. The corresponding quantum numbers are angular momentum and isospin,
projections of these quantum numbers, and parity. The second invariance is
invariance of the system in respect of identical particle permutations. For
the system composed of identical fermions, this symmetry is known as the Pauli
principle. The latter kind of correlations is created by the mentioned above
translational invariance. At any kind of correlations taken into account, one
gains some quantum numbers for the entire system, but at the same time
particles lose individuality and no one of them can be described by a complete
set of necessary quantum numbers. For example, binding the momenta one loses
momentum projections for individual particles. Antisymmetrizing the function,
the determinants appear, states for individual particles are not defined at
all. Only configuration, i.e., a set of restricted quantum numbers
characterizing the entire system, is defined. For translationally invariant
function one cannot define even the configuration.

Some recent methods, developed for few-body systems with strong interaction,
are designed to produce translationally invariant wave functions. These are
methods based on the Faddeev or Faddeev-Jakubovsky equations \cite{1}, on the
Green Function Monte Carlo (GFMC) method \cite{2}\ or on the expansions in a
large basis of harmonic oscillator functions \cite{3}. In the first case, the
translational invariance of wave function arises from equations written in
intrinsic variables. For GFMC the translational invariance is given by
infrastructure of variational wave function, spatial part of which is composed
of Jastrow multipliers dependent on translationally invariant differences of
one-particle coordinates. The mentioned latter calculations operate with the
harmonic oscillator wave functions as basis for realistic calculations. These
functions, if complete basis of ones corresponding to a given number of
oscillator quanta is taken into account, can be projected to superpositions
with unexcited center of mass of the system.

The goal of the current study is new efficient method of the construction of
the translationally invariant wave functions for the system of identical
fermions with arbitrary potential of pair interaction. These functions are
eigenfunctions of intrinsic Hamiltonian, i.e., Hamiltonian dependent on
intrinsic variables. One ensures translational invariance for wave functions
taking as arguments the set of translationally invariant spatial variables, so
called Jacobian variables, defined as in \cite{4}. The formalism is free of
any phenomenological parameters (as variational parameters in GFMC) or fields
(as in any model exploiting the shell model idea) as well as of any "effective interactions".

The basic point of the method is the well-known result that the essential
dynamic part of Hamiltonian for the system of identical particles is the
Reduced Hamiltonian (RH) operator \cite{5,6}. The system of eigenfunctions of
this operator takes into account all correlations caused by interaction. The
Schr\"{o}dinger equation for RH operator is very simple, not more complex than
the Schr\"{o}dinger equation for relative movement of two particles
interacting by given potential. This really looks like a one-particle problem,
like in the Shell Model. These eigenfunctions are basic in our formalism to
the wave function construction. The main problem in this approach appears to
be antisymmetrization of a component (i.e., the construction with lower degree
of antisymmetry) of wave function. The developed universal algorithm for
antisymmetrization gives us a possibility to perform this operation in a
simple way and keep numerical approximations under control. In the mentioned
basis with correlations taken into account in advance, convergence is fast.

In Sec. 2 we present the basic definitions and introduce the problem under
consideration. Sec. 3 is devoted to description of the new original procedure
for antisymmetrization of \ wave functions for translationally invariant
systems composed of identical fermions. In Sec. 4 one considers the entire
problem based on factorization of the Hamiltonian matrix. The summary of the
developed method, numerical test and conclusions are given in Sec. 5.

\section{Correlated components of wave function}

The suggested consideration is based on the presentation of an intrinsic
Hamiltonian for the system of $N$ identical particles as a sum of $N(N-1)/2$
two-particle operators:%

\begin{align}
\mathcal{H}  &  =-\frac{\hbar^{2}}{2m}\sum_{i=1}^{N}\nabla_{i}^{2}+\frac
{\hbar^{2}}{2mN}\nabla_{0}^{2}+\sum_{i,j=1\left(  i<j\right)  }^{N}V\left(
\mathbf{r}_{i}-\mathbf{r}_{j},\sigma_{i}\sigma_{j}\right) \nonumber\\
&  =\sum_{i,j=1\left(  i<j\right)  }^{N}\left[  -\frac{\hbar^{2}}{2mN}\left(
\nabla_{i}-\nabla_{j}\right)  ^{2}+V\left(  \mathbf{r}_{i}-\mathbf{r}%
_{j},\sigma_{i}\sigma_{j}\right)  \right]  \equiv\sum_{i,j=1\left(
i<j\right)  }^{N}h_{i,j},
\end{align}
where $V\left(  \mathbf{r}_{i}-\mathbf{r}_{j},\sigma_{i}\sigma_{j}\right)  $
is the potential of interaction, dependent, as usual, on difference of
radius-vectors of interacting particles $i$ and $j$ and on discrete degrees of
freedom $\sigma_{i}$, $\sigma_{j}$ of these particles, such as spins,
isospins, etc. The term $\hbar^{2}\nabla_{0}^{2}$\ $/2mN$ is the kinetic
energy operator of the center of mass of the system.

The arbitrary eigenvalue of this Hamiltonian, due to antisymmetry of
corresponding eigenfunction, equals the number of terms present in the sum,
multiplied by the expectation value of one of these operators, i.e.,%

\begin{equation}
\mathcal{E}=\left\langle \mathcal{H}\right\rangle =\left(
\begin{array}
[c]{c}%
N\\
2
\end{array}
\right)  \left\langle -\frac{\hbar^{2}}{mN}\nabla_{\mathbf{\xi}_{i,j}}%
^{2}+V\left(  \sqrt{2}\mathbf{\xi}_{i,j},\sigma_{i}\sigma_{j}\right)
\right\rangle \equiv\left(
\begin{array}
[c]{c}%
N\\
2
\end{array}
\right)  \left\langle h_{i,j}\right\rangle \equiv\left\langle H_{i,j}%
\right\rangle .
\end{equation}
Here $\left(
\begin{array}
[c]{c}%
N\\
2
\end{array}
\right)  =N\left(  N-1\right)  /2$ is a binomial coefficient and $\mathbf{\xi
}_{i,j}=\frac{1}{\sqrt{2}}\left(  \mathbf{r}_{i}-\mathbf{r}_{j}\right)  $ is
one of the Jacobian variables. To simplify the expressions, let us take
defined values for indices, namely $i=N-1$ and $j=N$ and mark the
corresponding Jacobian variable as $\mathbf{\xi\equiv\xi}_{N-1,N}$. The
operator $H_{N-1,N}\equiv H$ \ is called the\emph{\ }Reduced Hamiltonian
operator and is the main in further consideration. The best way to eliminate
dimensions of physical variables in above expressions is to present the
multiplier $\hbar^{2}/m$ in the form%

\begin{equation}
\frac{\hbar^{2}}{m}=\hbar\omega\times b^{2},
\end{equation}
where $\hbar\omega$ can be considered as a parameter of energy and $b$ as a
length parameter. Really, one free parameter appears here, $\omega$ or $b.$
Let us consider below the dimension-free values for energy $\mathcal{E=}$
$\mathcal{E}/\hbar\omega$\ and variable $\mathbf{\xi}=$ $\mathbf{\xi}/b.$ The
value for parameter $b$ without any problems can be taken equal to one.
Finally, the dimension-free Reduced Hamiltonian equals%

\begin{equation}
H=\left(
\begin{array}
[c]{c}%
N\\
2
\end{array}
\right)  \left[  -\frac{1}{N}\Delta_{\mathbf{\xi}}+\frac{1}{\hbar\omega
}V\left(  \sqrt{2}\mathbf{\xi},\sigma_{N-1}\sigma_{N}\right)  \right]  .
\end{equation}

Thus, the expression for the eigenvalue of Hamiltonian in terms of
\ $\hbar\omega$ is $\mathcal{E}=\left\langle H\right\rangle .$

This is a very interesting result stating that calculation of eigenvalue does
not require the total Hamilton operator. Only the significant part of the
many-particle Hamiltonian with pair interaction - the two-particle RH operator
- is necessary. However, practical application of simplifications caused by
this observation is not easy. Due to noncommutation of $H$ and\ $\mathcal{H}%
$\ operators and antisymmetry of wave function one cannot prescribe some set
of quantum numbers for the state of the last two particles. The way of solving
this problem was found years ago, when coefficients of fractional parentage
(CFP) for the atomic shell model were introduced \cite{7,8}. These are the
coefficients of the antisymmetric wave function expansion in terms of linear
combinations of some easier structures with a lower degree of antisymmetry,
later on called components of wave function. For the problem under
consideration the components are products of antisymmetric function for the
last two particles dependent on variables present in the given Reduced
Hamiltonian operator, and the antisymmetric function dependent on a set of all
residual variables \cite{6}. In other words, the idea of coefficients of
fractional parentage realizes the antisymmetrization in space of quantum
numbers of the complete system of components. Let us mark the antisymmetric
wave function for the $N$ particle system as%

\begin{equation}
\Psi_{\mathcal{E}\Lambda M}\left(  1,2,...,N-1,N\right)  ,
\end{equation}
where $\mathcal{E}$\ is the corresponding eigenvalue of Hamiltonian, $\Lambda$
is a set of "good" quantum numbers, such as momentum, parity, isospin, etc.,
and $M$ is a set of corresponding projections of momenta quantum numbers.

The mentioned construction with a lower degree of antisymmetry, i.e., the
component of wave function, is:%

\begin{equation}
\Phi_{\left(  \bar{\Gamma}\bar{\Lambda},\varepsilon\rho\right)  \Lambda
M}\left(  1,2,...;N-1,N\right)  =\sum_{\bar{M}\mu}\Psi_{\bar{\Gamma}%
\bar{\Lambda}\bar{M}}\left(  1,2,...,N-2\right)  \varphi_{\varepsilon\rho\mu
}\left(  N-1,N\right)  \left[
\begin{array}
[c]{ccc}%
\bar{\Lambda} & \rho & \Lambda\\
\bar{M} & \mu & M
\end{array}
\right]  .
\end{equation}
It is a bound momenta function with exact quantum numbers $\Lambda M$
coinciding with ones for complete wave function. The function $\Psi$ present
on the right-hand side of equation is a \textquotedblright
spectator\textquotedblright\ function dependent on variables of first $\left(
N-2\right)  $ particles. $\bar{\Lambda}\bar{M}$ is a set of corresponding
quantum numbers and its projections. $\bar{\Gamma}$\ marks a set of all the
remaining quantum numbers necessary to ensure completeness and
orthonormalization of basis of "spectator" functions. The second function
$\varphi$\ is eigenfunction of the RH operator, hence the best way is to
choose the quantum numbers $\rho$ and $\mu$ as a complete set of eigenvalues
of operators commuting with the RH operator. $\varepsilon$\ marks an
eigenvalue of the Schr\"{o}dinger equation for RH. The last factor is the
product of Clebsh-Gordan coefficients for momenta and Kronecker deltas $-$ for
parities, isospin projections, etc.

The most interesting for us are coefficients of the wave function expression
in terms of a complete set of components:%

\begin{equation}
\Psi_{\mathcal{E}\Lambda M}\left(  1,2,...,N-1,N\right)  \newline=\sum
_{\bar{\Gamma}\bar{\Lambda},\varepsilon\rho}\left\langle \mathcal{E}%
\Lambda\Vert\bar{\Gamma}\bar{\Lambda},\varepsilon\rho\right\rangle
\Phi_{\left(  \bar{\Gamma}\bar{\Lambda},\varepsilon\rho\right)  \Lambda
M}\left(  1,2,...;N-1,N\right)  .
\end{equation}
The system of equations for these coefficients consists of the Schr\"{o}dinger
equation and condition that wave function of the system, composed of identical
fermions, must be an eigenfunction of the antisymmetrization operator
corresponding to the eigenvalue equal to one:%

\begin{equation}
\left(  \mathcal{H-E}\right)  \Psi_{\mathcal{E}\Lambda M}\left(
1,2,...,N-1,N\right)  =0,
\end{equation}

\begin{equation}
\left(  A\mathcal{-}1\right)  \Psi_{\mathcal{E}\Lambda M}\left(
1,2,...,N-1,N\right)  =0. \label{devinta}%
\end{equation}

Obviously, the antisymmetrizer, present in Eq. (\ref{devinta}), must be
properly normalized, i.e.:%

\begin{equation}
A\mathcal{=}\frac{1}{N!}\sum_{P\in S_{N}}\delta_{P}P, \label{antra}%
\end{equation}
where $P$\ are permutation operators of symmetric group $S_{N}$\ and
$\delta_{P}$ is the parity of permutation $P$. Namely due to this
normalization the antisymmetrizer is a projection operator, i.e., satisfies
the condition $AA\mathcal{=}A.$ The same condition for matrix of this operator
simplifies further consideration.

However, simultaneous solution of both equations is very problematic. As usual
in quantum mechanics, first of all one needs to construct a complete set of
solutions for a simpler equation and later on, applying these functions, to
construct solutions of a more complex equation. The simpler equation in our
case is the second equation, so first of all it is necessary to antisymmetrize
the components. The antisymmetrized basic functions, eigenfunctions of
antisymmetrizer, are:%

\begin{equation}
\Psi_{\Gamma\Lambda M}\left(  1,2,...,N-1,N\right)  =\sum_{\bar{\Gamma}%
\bar{\Lambda},\varepsilon\rho}\left\langle \Gamma\Lambda\Vert\bar{\Gamma}%
\bar{\Lambda},\varepsilon\rho\right\rangle \Phi_{\left(  \bar{\Gamma}%
\bar{\Lambda},\varepsilon\rho\right)  \Lambda M}\left(  1,2,...;N-1,N\right)
.
\end{equation}
Namely coefficients, present in this expansion, are the coefficients of
fractional parentage; $\Gamma$\ is an additional quantum number necessary to
mark the antisymmetric basic functions. The last operation is diagonalization
of the Hamiltonian operator while applying this basis of functions. For this
one needs to define the expansion of wave function%

\begin{equation}
\Psi_{\mathcal{E}\Lambda M}\left(  1,2,...,N-1,N\right)  =\sum_{\Gamma
}\left\langle \mathcal{E}\mid\Gamma\right\rangle \Psi_{\Gamma\Lambda M}\left(
1,2,...,N-1,N\right)
\end{equation}
and to solve an algebraic eigenvalue equation for these coefficients:%

\begin{equation}
\sum_{\Gamma^{^{\prime}}}\left[  \left\langle \Gamma\left\vert \mathcal{H}%
\right\vert \Gamma^{\prime}\right\rangle -\mathcal{E}\right]  \left\langle
\mathcal{E}\mid\Gamma^{\prime}\right\rangle =0.
\end{equation}

The final expression for wave function in terms of components with a lower
degree of antisymmetry looks as:%

\begin{equation}
\Psi_{\mathcal{E}\Lambda M}\left(  1,2,...,N-1,N\right)  =\sum_{\Gamma
}\left\langle \mathcal{E}\mid\Gamma\right\rangle \sum_{\bar{\Gamma}%
\bar{\Lambda},\varepsilon\rho}\left\langle \Gamma\Lambda\Vert\bar{\Gamma}%
\bar{\Lambda},\varepsilon\rho\right\rangle \Phi_{\left(  \bar{\Gamma}%
\bar{\Lambda},\varepsilon\rho\right)  \Lambda M}\left(  1,2,...;N-1,N\right)
.
\end{equation}

\section{Antisymmetrization in correlated basis}

Let us first of all formulate the well-known results, concerning
antisymmetrization of different wave functions in a way giving possibility to
generalize it for translationally invariant functions.

The best known procedure is antisymmetrization of many-particle function
dependent on one-particle spatial variables and intrinsic degrees of freedom,
when particles do not interact, correlations are absent, and wave functions
split into product of one-particle functions. In this case, antisymmetrization
can be performed in given configuration, i.e., for each set of one-particle
states provided that coinciding one-particle states are absent. This
construction equals the determinant composed of mentioned one-particle
functions with permuted one-particle variables.

The simplest fractional parentage expansion for this wave function is
expansion for determinant in minors, i.e., in constructions antisymmetric only
in respect of permutations of the first $\left(  N-1\right)  $\ particles. The
last particle is not antisymmetrized with particles composing this "spectator"
system. However, a state for this particle in an antisymmetrized function runs
over all states present in configuration under investigation. The coefficients
of expansion for antisymmetric function, equal to the determinant of the
$N$-th order in terms of minors of $\left(  N-1\right)  $ order, are the
one-particle CFP. They are well-known from linear algebra. Due to
normalization of wave function and all minors they equal $\pm1/\sqrt{N}.$ In
this case, definition for the two-particle CFP is completely analogous. They
are coefficients present in the same determinant expansion in terms of minors
of the $\left(  N-2\right)  $-th order multiplied by antisymmetric wave
function for remaining two particles.

The antisymmetrization and fractional parentage expansion for more complex
function with bound momentum-like, i.e., orbital, spin, isospin, etc., quantum
numbers, can be formulated in analogous way. In this case, there appear
shells, i.e., groups of particles having the same one-particle radial wave
function but different projections of orbital and intrinsic momenta of
particles. However, the antisymmetrization for such a function as in previous
case is impossible. The operation of momenta binding results in sums over
projections of quantum numbers, hence a complete set of one-particle variables
for every particle stays undefined and the old scheme of antisymmetrization
cannot work. Namely the antisymmetrization in the defined shell creates
traditional coefficients of fractional parentage. As in the previous case,
they are independent of one-particle radial wave functions, hence are
universal. For more complex systems, containing more than one shell,
antisymmetrization is not so complicated as believed because radial functions
for different shells are orthogonal, and this significantly simplifies
consideration. The CFP, defined for such a complex configuration, are
expressible in terms of traditional CFP and transformation matrices. Again in
this case antisymmetrization can be performed for defined configuration, it is
in finite space of components basis.

Antisymmetrization for translationally invariant wave function is a very
complex problem in comparison with considered ones. In the form given below,
it is introduced in \cite{6}\ and still applied only to three-dimensional
harmonic oscillator functions \cite{3}. This is the simplest application of
the proposed scheme because only for harmonic oscillator basis this operation
can be performed in finite space of basic component functions. The reason for
this is a very simple structure of the many-particle harmonic oscillator
Hamiltonian, invariant in respect of any permutations of Jacobian variables.
This symmetry creates a conserving quantum number - a total number of
oscillator quanta proportional to the eigenvalue of such a Hamiltonian, and
the possibility to perform antisymmetrization separately for each value of
this integer number. However, this simplification is not so useful as it looks
at first glance because in many cases long expansions in this basis are
necessary and convergence of expansions is very slow.

The procedure one applies is defined in a universal way and is based on
calculation of the antisymmetrizer matrix in components basis and spectral
decomposition of this matrix in case when the eigenfunctions of the RH
operator are applied as "actor" functions. The formalism is devoted to bound
states of the quantum system description, so in expansion of component one can
apply square-integrable functions for "spectator", as well as
square-integrable functions for "actor". Obviously, in most interesting
applications RH has both discrete and continuous eigenvalues. The necessary
basis for the corresponding Sturm-Liouville problem can be constructed taking
proper boundary conditions for eigenfunctions, for example, taking all
eigenfunctions equal to zero at some value of argument larger in comparison
with a characteristic measure for the system dimensions.

The next problem is a complete system of functions for "spectator". For these
functions one cannot formulate any equations, hence one is free in choice. The
simplest translationally invariant basis, as mentioned above, is basis of
three-dimensional harmonic oscillator functions, enriched by intrinsic degrees
of freedom of particles.

Once the construction of basis for components is completed, one can start
calculating the matrix of an antisymmetrization operator. If, as defined, one
has antisymmetrized "spectator" as well as "actor" functions and is going to
apply this operator to the matrix calculation, the best factorization for the
antisymmetrization operator is as follows \cite{9}:%

\begin{equation}
A_{1,2,...,N}=A_{N-1,N}A_{1,2,...,N-2}Y_{1,2,...,N-2;N-1,N}A_{1,2,...,N-2}%
A_{N-1,N}, \label{trecia}%
\end{equation}
where%

\begin{equation}
Y_{1,2,...,N-2;N-1,N}=\left(
\begin{array}
[c]{c}%
N\\
2
\end{array}
\right)  ^{-1}[1-\left(  N-2\right)  P_{N-2,N}+\left(
\begin{array}
[c]{c}%
N-2\\
2
\end{array}
\right)  P_{N-3,N-1}P_{N-2,N}] \label{ketvirta}%
\end{equation}
and $P_{i,j}$ are \ transposition operators.

The normalization of antisymmetrizer given in Eq.(\ref{antra}) makes
corresponding matrix idempotent or projection matrix:%

\begin{equation}
\mathbf{AA}=\mathbf{A}.
\end{equation}

This matrix is real and symmetric $\left(  \mathbf{A}^{+}=\mathbf{A}\right)
$. Obviously, its eigenvalues equal ones and zeroes. Hence, the sum of
eigenvalues of this matrix equals the number of eigenvalues equal one. By
definition, it coincides with trace of matrix $\mathbf{A}$ and defines the
rank $r$ of this matrix. Eigenvectors, corresponding to zero eigenvalues, do
not have a physical meaning. Antisymmetric states can be described only by
eigenvectors corresponding to eigenvalue equal one. This matrix has an
attractive feature\ - each entry $a_{ij}$\ of this matrix equals the product
of the $i$-th and $j$-th rows (columns) of this matrix. Hence, each diagonal
entry of matrix equals the sum of squares of entries of this row or column.
This is a useful measure of projectiveness of any calculated part of matrix.
Moreover, if some diagonal entry of this matrix equals zero, all entries of
corresponding row and column are identical zeroes. This means that such a
basic component must be removed from basis because it has some additional
symmetries and cannot be antisymmetrized. The second conclusion from this
feature of the projection matrix is that if some diagonal entry of this matrix
equals one, again all remaining entries of corresponding row and column equal
zero. This means that the corresponding basic component is already
antisymmetric and needs no additional antisymmetrization.

The spectral decomposition for this matrix is%

\begin{equation}
\mathbf{A}=\mathbf{FF}^{+},
\end{equation}
where $\mathbf{F}$\ is the matrix composed of eigenvectors, corresponding to
unit eigenvalues, as columns. Every column of this matrix defines
antisymmetric wave function for $N$ particles. Matrix $\mathbf{F}$\ has $n$
rows ($n$ equals order of matrix $\mathbf{A}$) and $r$ columns. Every column
is normalized and they are orthogonal, i.e.:%

\begin{equation}
\mathbf{F}^{+}\mathbf{F}=\mathbf{1}.
\end{equation}

The entries of matrix $\mathbf{F}$\ are namely the coefficients of fractional
parentage for any of mentioned components bases. For this matrix construction,
observation that every column of matrix $\mathbf{A}$ is eigenvector of this
matrix, corresponding to the eigenvalue equal one, is very useful. However,
the number of these eigenvectors exceeds the rank, hence some of them are
linearly-dependent. Moreover, they are not orthonormal vectors, but direct
calculation of matrix $\mathbf{F}$ is not the problem due to the well-known
result that each symmetric matrix of an order, equal $n,$ and a rank, equal
$r$, can be written in a form%

\begin{equation}
\mathbf{A}=\left(
\begin{array}
[c]{ccc}%
\mathbf{Z} & \  & \mathbf{Q}^{+}\\
\mathbf{Q} & \  & \mathbf{QZ}^{-1}\mathbf{Q}^{+}%
\end{array}
\right)  , \label{penkta}%
\end{equation}
where $\mathbf{Z}$\ is nondegenerate submatrix of the order $r$. The matrix of
spectral decomposition for matrix $\mathbf{A}$ is:%

\begin{equation}
\mathbf{F}=\left(
\begin{array}
[c]{c}%
\mathbf{Z}^{1/2}\\
\mathbf{QZ}^{-1/2}%
\end{array}
\right)  . \label{sesta}%
\end{equation}
The normalization condition for $\mathbf{F}$ looks as%

\begin{equation}
\mathbf{F}^{+}\mathbf{F}=\mathbf{Z}^{+}\mathbf{Z}^{-1/2}\mathbf{Q}%
^{+}\mathbf{QZ}^{-1/2}=\mathbf{1.}%
\end{equation}
It can be written in the form%

\begin{equation}
\mathbf{Q}^{+}\mathbf{Q}=\mathbf{Z}\left(  \mathbf{1}-\mathbf{Z}\right)  .
\end{equation}

\section{Eigenvalues of Hamiltonian matrix}

Once the matrix $\mathbf{F}$ is constructed, one is ready to obtain a matrix
of the total Hamiltonian. As it was mentioned, first of all one must take into
account that this matrix due to identity of particles equals the matrix of the
RH operator. Having in mind that the coefficients of the antisymmetric
function expression in terms of components are CFP, i.e., entries of
$\mathbf{F}$ matrix, one can present the matrix of the Hamiltonian in a form%

\begin{equation}
\mathcal{H}\mathbf{=F}^{+}\mathbf{HF}.
\end{equation}
Here the matrix of the RH operator is given in components basis. This is our
goal because in this basis matrix $\mathbf{H}$ is diagonal. The best choice is
to rearrange the entries of this matrix in the nonvanishing order. By the way,
this ordering of basic functions looks very natural and means that one takes
the function with "actor" being in the lowest possible state as the first
basic function, then in the first excited state - as the second basic
function, etc. Obviously, for every "actor" function one has to take a
necessary set of "spectator" functions.

The order of the Hamiltonian matrix equals rank $r$ of matrix of an
antisymmetrizer, but the order of the diagonal matrix of the RH operator is
larger, equal to the order of the antisymmetrizer matrix $n$\ or, in other
words, the number of basic components taken into account. This infrastructure
of the Hamiltonian matrix is very useful in cases when dimensions of matrices
present in expression are very large. Taking into account this infrastructure,
one can construct different approximations for eigenvalues and eigenfunctions.
The idea is as follows. The antisymmetric states, involved in the expansion,
are given by columns of matrix $\mathbf{F} $.\ Let the rank of this matrix be
equal one, this means it contains only one column. In this case the eigenvalue
of the Hamiltonian equals:%

\begin{equation}
\mathcal{E}=\varepsilon_{1}f_{11}^{2}+\varepsilon_{2}f_{21}^{2}%
+...+\varepsilon_{n}f_{n1}^{2}\equiv\varepsilon_{1}\omega_{1}+\varepsilon
_{2}\omega_{2}+...+\varepsilon_{n}\omega_{n}.
\end{equation}
Here $\varepsilon_{i}$ are the eigenvalues of the RH operator arranged in the
nonvanishing order. Due to reality of matrix $\mathbf{F}$\ the squares of
entries are the probabilities (due to normalization this sum of squares equals
one, i.e., $\sum_{i=1}^{n}f_{i1}^{2}=1$) of different states of the RH
operator in the expansion for the energy eigenvalue of the total Hamiltonian.
The expansion of the same kind can be obtained for each eigenvalue after
matrix diagonalization because the corresponding eigenvector equals a linear
combination of columns of matrix $\mathbf{F}$:%

\begin{equation}
\mathcal{E}=\sum_{i=1}^{n}\varepsilon_{i}\omega_{i},\quad\sum_{i=1}^{n}%
\omega_{i}=1,\quad\omega_{i}\geq0. \label{septinta}%
\end{equation}

As it follows from definitions, all $\omega_{i}$\ are nonnegative. At the same
time some first eigenvalues of RH are negative and remaining ones $-$
positive. To minimize the value for energy one needs values of probabilities,
corresponding to negative values of epsilons, as maximal as possible.\ Namely
the possibility to obtain this formula for the eigenvalue of the many-particle
Hamiltonian is basic to the Hall-Post lower bounds for energies construction
\cite{5}. This lower bound can be obtained taking a maximal value of
probability corresponding to the lowest value of RH energy. This means:%

\begin{equation}
E_{Hall-Post}=\varepsilon_{1}.
\end{equation}
Obviously, this is namely a lower bound because%

\begin{equation}
E-E_{Hall-Post}=\sum_{i=1}^{n}\varepsilon_{i}\omega_{i}-\varepsilon_{1}%
\sum_{i=1}^{n}\omega_{i}=\sum_{i=2}^{n}\left(  \varepsilon_{i}-\varepsilon
_{1}\right)  \omega_{i}>0.
\end{equation}
On the other hand, our consideration has shown that this lower bound value for
energy corresponds to wave function equal to the first basic component. This
component is not antisymmetric, hence the corresponding eigenvalue is
significantly lower than that for the antisymmetrized function.

The real problem in our task is a huge amount of basic states while describing
strongly correlated systems and as consequence $-$ a very big rank of matrix
of the antisymmetrizer, hence a large order of the Hamiltonian matrix. One
needs the best possible values for energy and the high quality corresponding
wave functions at the lowest possible price. The recommendations for
successful realization of this program are as follows.

First of all, one does not need the complete calculation of matrix of the
antisymmetrizer $\mathbf{A}$. As one may conclude from this matrix
presentation, Eq.(\ref{penkta}), for complete matrix and its spectral
decomposition construction only $r$ linearly independent columns or rows of
this matrix, i.e., matrices $\mathbf{Z}$ and $\mathbf{Q}$, are necessary.
These define the matrix of spectral decomposition $\mathbf{F}$,
Eq.(\ref{sesta}),\ necessary for further calculations. During calculations of
this part of matrix $\mathbf{A}$ one can control how far this calculation is
from finish because due to the mentioned above condition the sum of squares of
entries of column must equal the corresponding diagonal entry. When this
condition is fulfilled with necessary precision, one can stop calculations for
this column and start calculating the next column. The corresponding submatrix
$\mathbf{Z}$ will show how many columns one needs. As it is required, this
matrix must be nondegenerate, hence its determinant must be nonzero. Only in
this case, one can continue column calculations. If the determinant equals
zero, the basic state, corresponding to column under consideration, is
linearly dependent on already calculated columns. This statement follows from
relation for entries of projection matrix, stating that each submatrix of the
projection matrix equals the Sturm matrix of corresponding rows or columns.
Hence, the determinant of submatrix $\mathbf{Z}$\ is the best indicator of the
linear dependence of rows (columns) of projection matrix. Finally, one can
check the rank of matrix. It must equal the trace.

The last operation, when both matrices, $\mathbf{F}$ and $\mathbf{H,}$ are
calculated, is construction of the Hamiltonian matrix and its diagonalization.
This is the standard operation. However, in many applications only a few
lowest eigenvalues and eigenvectors of this matrix are necessary. The method
of obtaining converging approximations to exact values for a few lower states
avoiding calculation of the complete Hamiltonian matrix is developed. The idea
is based on applied transparent presentation for the Hamiltonian. As usual,
the diagonalization of this matrix requires the set of orthogonal transformations%

\begin{equation}
\mathbf{R}^{+}\mathbf{F}^{+}\mathbf{HFR=}\left(  \mathbf{FR}\right)
^{+}\mathbf{HFR.}%
\end{equation}
In our approach the transformation for the entire matrix is not necessary -
one can transform the matrix $\mathbf{F}$ in a necessary way. The needed
transformations are products of $(r-1)$ Householder reflections \cite{10},
i.e., $\mathbf{R=R}_{1}\mathbf{R}_{2}...\mathbf{R}_{r-1}$ \ transforming the
first, second and all the following rows of matrix $\mathbf{F}$ to convert it
into the lower triangle form. To some extent this way of matrix transforming
looks like the well-known QR decomposition of matrix. If the first row of
matrix $\mathbf{F}$ is $\mathbf{f}_{1}^{+}=\left(  f_{11}f_{12}...f_{1r}%
\right)  ,$ the first Householder matrix, whose order equals $r,$ is%

\begin{equation}
\mathbf{R}_{1}=\mathbf{1-}2\frac{\mathbf{f}_{1}\mathbf{f}_{1}^{+}}%
{\mathbf{f}_{1}^{+}\mathbf{f}_{1}}.
\end{equation}

The second Householder matrix is%

\begin{equation}
\mathbf{R}_{2}=\left(
\begin{array}
[c]{ccc}%
\mathbf{1}_{1} & \  & \mathbf{0}^{+}\\
\mathbf{0} & \  & \mathbf{1-}2\frac{\mathbf{f}_{2}\mathbf{f}_{2}^{+}%
}{\mathbf{f}_{2}^{+}\mathbf{f}_{2}}%
\end{array}
\right)  ,
\end{equation}
where $\mathbf{f}_{2}^{+}=\left(  f_{22}f_{23}...f_{2r}\right)  $ is the upper
(right) part of the second row of matrix $\mathbf{F}$ and the order of unit
submatrix is shown. The last matrix of transformation equals:%

\begin{equation}
\mathbf{R}_{r-1}=\left(
\begin{array}
[c]{ccc}%
\mathbf{1}_{r-2} & \  & \mathbf{0}^{+}\\
\mathbf{0} & \  & \mathbf{1-}2\frac{\mathbf{f}_{r-1}\mathbf{f}_{r-1}^{+}%
}{\mathbf{f}_{r-1}^{+}\mathbf{f}_{r-1}}%
\end{array}
\right)  .
\end{equation}
Here $\mathbf{f}_{r-1}^{+}=\left(  f_{r-1r}f_{rr}\right)  .$ After these
transformations are performed, the new Hamiltonian matrix $\mathbf{F}%
^{\prime+}\mathbf{HF}^{\prime}$\ appears, where $\mathbf{F}^{\prime
}=\ \mathbf{FR}_{1}\mathbf{R}_{2}...\mathbf{R}_{r-1}.$ Obviously, the new
matrix has the same eigenvalues as the initial one. Moreover, $\mathbf{F}%
^{\prime+}\mathbf{F}^{\prime}=\mathbf{1}$ and $\mathbf{F}^{\prime}%
\mathbf{F}^{\prime+}=$ $\mathbf{A}$. The matrix $\mathbf{F}^{\prime}$ has
zeros in the upper triangle and maximal possible diagonal entries, equal to
norms of parts of entries of rows, involved in the Householder reflection,
i.e., $\mathbf{f}_{1}^{+}\mathbf{f}_{1},\mathbf{f}_{2}^{+}\mathbf{f}_{2},$...,
$\mathbf{f}_{r-1}^{+}\mathbf{f}_{r-1},$ respectively.

Now one can obtain that nonzero entry in the first row has only the first
column of matrix $\mathbf{F}^{\prime}.$ This entry squared equals the maximal
weight (probability) of the lowest eigenvalue of RH. Any combinations of this
column with remaining columns of matrix $\mathbf{F}^{\prime}$ can only lower
this probability. The nonzero weight (probability) of the first and second of
the lowest eigenvalues of RH can produce only the first and second columns,
etc. The set of negative eigenvalues of RH, as mentioned, is finite. Let this
number be equal $k$. Not more than $\left(  k-1\right)  $ Householder
reflections\ are necessary to obtain an acceptable result because, as already
mentioned, each column with zero entries, corresponding to negative RH
eigenvalues, while combining with the first $k$ columns, can only lower the
probabilities of negative epsilons, hence a negative part of the eigenvalue
for the total Hamiltonian.

\section{Summary and conclusions}

Summarizing, let us present the list of consequent steps for the application
of the proposed method.

The first step is construction of the Reduced Hamiltonian operator given in
Eq.(4), and generation of the complete system of square-integrable functions
for this operator. The spectrum of this Sturm-Liouville problem can be
discretized by applying simple boundary conditions for eigenfunctions.

The next step is construction of the system of basic functions for components,
taking arbitrary antisymmetric functions for "spectator" and eigenfunctions of
RH - for "actor" basis. The components must be arranged in the order
corresponding to nonvanishing diagonal entries of matrix $\mathbf{H,}$ i.e.,
eigenvalues of the RH operator.

Now one can start calculation of the antisymmetrizer matrix using
factorization, given in Eq.(15). This presentation is optimal for component
basis because due to antisymmetry of "spectator" and "actor" functions both
antisymmetrizers present on the left-hand side as well as both
antisymmetrizers, present on the right-hand side, may be omitted. Obviously,
one must calculate only matrices of two nontrivial operators - $P_{N-2,N}%
$\ and $P_{N-3,N-1}P_{N-2,N}$. Later on, combining them with the identity
matrix according to Eq.(16), one can construct the matrix of the
antisymmetrizer. Due to simple structure of this matrix, thoroughly considered
in Sec. 3,\ one needs only a part of this matrix, necessary for matrix
$\mathbf{F}$ construction according to formulas, given in Eqs. (20) and (21).
By the way, one can start calculations of eigenvalues for the total
Hamiltonian at any number of columns (rows) of the antisymmetrizer matrix,
once it is not less than the number $k$ of negative eigenvalues of the RH
operator. The complete calculation requires $r$ columns of matrix $\mathbf{A}%
$, but if one stops at some other number of columns calculated, this is more
or less successful approximation for the rank of this matrix.

The last step is application of the first, second, third and following
Householder reflections and diagonalizations of corresponding matrices, whose
orders equal the number of reflection. Due to the variational character of
these approximations, the results obtained, as usual, will converge to precise
values from above.

The application of the method for noninteracting particles is the simplest and
most transparent. In this situation, RH is proportional to the Hamiltonian,
characteristic of any Shell Model picture. It is the Hamiltonian for one
particle, moving in the external field, multiplied by the number of particles
$N$, hence its eigenvalues are well-known one-particle energies, multiplied by
the same number. As it was mentioned, the matrix of the antisymmetrizer now is
independent of any dynamics, and it splits down into a direct sum of
projection matrices for different configurations. One is free to take spectral
decompositions for every of these matrices separately. If particles are
without any correlations, i.e., if the system can be described by one
determinant, probabilities of all one-particle states in given configuration
equal $1/N$, hence the sum has a well-known form given in Eq.(26). In case
when one must ensure good quantum numbers for such a system, the probabilities
of different states of RH will equal the squared coefficients of fractional
parentage, obtained after spectral decomposition of the antisymmetrizer matrix.

Let us take for illustration of proposed ideas a simple model for $^{1,3}F$
terms in $d^{4}$ configuration of four identical fermions with spin equal 1/2.
The order $n$ of this matrix equals $8$, and the rank $r$ equals $3$. The
matrix $\mathbf{F}$ of spectral decomposition is:%

\begin{equation}
\mathbf{F}^{+}=\left(
\begin{array}
[c]{cccccccc}%
0 & \sqrt{3/14} & 0 & \sqrt{5/14} & 0 & -\sqrt{3/16} & -\sqrt{3/560} &
-\sqrt{33/140}\\
-\sqrt{1/15} & \sqrt{2/105} & \sqrt{3/8} & \sqrt{1/56} & \sqrt{4/15} &
-\sqrt{1/60} & \sqrt{3/28} & \sqrt{11/84}\\
-\sqrt{4/15} & -\sqrt{5/42} & 0 & \sqrt{1/14} & -\sqrt{1/15} & -\sqrt{5/48} &
-\sqrt{27/112} & \sqrt{11/84}%
\end{array}
\right)
\end{equation}

Here, to make the text easy, one presents the transposed matrix. Let diagonal
entries $\varepsilon_{i}$\ of matrix of RH be:%

\begin{equation}
-7;-4;0;0.1;0.2;0.3;0.4;0.5.
\end{equation}

These eigenvalues are not far from the real situation. Existence of a few
bound states of RH and all remaining states situated at positive energies,
i.e., in the continuous spectrum, is characteristic of strong interaction. As
mentioned, in our approach the continuous spectrum is discretized due to
boundary conditions chosen.

The first state (row) of $\mathbf{F}^{+}$\ (the first column of $\mathbf{F}$)
matrix produces a very pure value for energy, equal $-0.645,$ the second -
even worse, equal $-0.374$ due to the smaller total weight of basic states,
corresponding to the first two negative eigenvalues of RH. The result,
produced while applying the third state (row) of $\mathbf{F}^{+}$, is the best
one. It equals $-2.\,129.$ This value is closest to the lowest eigenvalue of
the total Hamiltonian matrix $\mathbf{F}^{+}\mathbf{HF}$, which equals
$-2.\,532\mathbf{.}$

The result of the first Householder reflection, the first row of $F^{\prime
+}=R_{1}^{+}F^{+},$\ gives the eigenvalue equal to\ $-2.\,380.$\ After the
second Householder reflection applied, the lowest eigenvalue of the second
order matrix equals $-2.\,532,$\ hence at a given precision it coincides with
the exact eigenvalue of the total matrix of the Hamiltonian.

The above mentioned Hall-Post lower bound for this eigenvalue equals $-7.000.
$ The diagonalization of the corresponding third order matrix in uncorrelated
harmonic oscillator basis due to the core present in realistic nucleon-nucleon
potential can give only positive values for the RH operator, hence a positive
value for energy of the whole system. Obviously, our result is a real
achievement for very complex interactions, such as recent realistic nucleon -
nucleon potentials.

Finally, let us draw some parallel between the developed approach and the
well-known Hartree-Fock Self-Consistent Field method. In both cases first of
all the building blocks for the system wave function construction are
necessary. The common feature in both approaches is construction of these
blocks taking into account as many dynamic correlations as possible. In SCF
these, due to the chosen scheme of simple antisymmetrization in given
configuration, are the best possible one-particle functions. In our approach,
when one can choose the refined way of antisymmetrization, these are the
eigenfunctions of the reduced Hamiltonian operator. Both approaches at this
stage have some attractive points. The SCF method applies determinants or CFP.
The price for this simplicity is well-known unavoidable appearance of the
phenomenological central field, the conversion of the linear problem to
nonlinear equation for one-particle wave functions, and the essential
impossibility to take into account some dynamic correlations while applying
this method. In our method, all dynamic correlations are taken into account in
advance by using RH eigenfunctions, however now one can perform
antisymmetrization only in infinite space. The next step in SCF is iterations
for the self-consistent field and optimization of corresponding one-particle
functions. In our approach, this corresponds to the application of $k$
Householder reflections and diagonalizations of the Hamiltonian matrix of the
growing order. This procedure is equivalent to the lowest configuration
definition in the SCF method. Finally, if SCF approach in some cases does not
work, superpositions of configurations are necessary. This essentially
complicates the consideration. In our case in analogous situation, one needs
to take more columns of the antisymmetrizer matrix, no more serious problems,
no new complex operations. Our approach can be applied to the translationally
invariant functions, while for the Self-Consistent Field approach such a
modification is impossible.

As a more realistic application let us comment on the modifications of
traditional calculation in the harmonic oscillator basis for the three nucleon
system \cite{11}. The number of basic states (components) with the number of
oscillator quanta up to E=44 is 16744. This equals the dimension of both - RH
and antisymmetrizer matrices. The rank of the antisymmetrizer matrix, equal to
the dimension of the total Hamiltonian matrix, is 5553. The ground state
energy of triton in this approximation for the realistic nucleon-nucleon
potential Reid93 \cite{12} equals -7.59 MeV (the exact value of ground state
energy for this potential is -7.63). Among 5553 basic components only 131 have
"actor" wave-functions, corresponding to negative eigenvalues of RH in two
nucleon channels $1^{+}0$ $\left(  ^{3}S_{1}-^{3}D_{1}\right)  $ and $0^{+}1$
$\left(  ^{1}S_{0}\right)  $. Therefore, applying 130 Householder reflections
for columns of matrix $\mathbf{F}$ before the total Hamiltonian matrix
calculation one can transform it to lower triangle form with only 131 columns,
having nonzero entries, corresponding to negative eigenvalues of the RH
operator. Namely these entries produce nonzero probabilities of corresponding
states. As a result, only these 131 basic antisymmetric functions (transformed
columns of matrix F) are necessary for construction of the total Hamiltonian
matrix, ensuring the mentioned result for the triton ground state energy.
Obviously, transformation of dimension of the Hamiltonian matrix from 5553 to
131 is a real achievement of the proposed method, giving opportunity to
simplify the realistic nuclear calculations.

\bigskip

\textit{D.G. and R.K. are grateful to the Lithuanian State Science and Studies
Foundation in the frame of project No T-12/06 for the partial support.}

\end{document}